# Interplanetary drivers of the magnetospheric disturbances: A brief review of incorrect approaches

**Yuri Yermolaev**[1], Irina Lodkina[1], Lidia Dremukhina[2], Michael Yermolaev[1], and Alexander Khokhlachev[1]

[1]Space Research Institute (IKI RAN), Moscow, Russian Federation (yermol@iki.rssi.ru)
[2]Pushkov Institute of Terrestrial Magnetism, Ionosphere, and Radio Wave Propagation (IZMIRAN), Troitsk, Moscow, Russian Federation

One of the most promising areas of research in solar-terrestrial physics is the comparison of the responses of the magnetosphere-ionosphere-atmosphere system to various interplanetary disturbances (the so-called "interplanetary dravers"). Numerous studies show that different types of drivers cause a different reaction of the system for identical IMF variations. At the same time, the number of incorrect approaches in this direction of research has increased. These errors can be attributed to 4 large classes. (1) The first class includes works whose authors uncritically reacted to previously published works with incorrect driver identification and use incorrect results in their work. (2) Some authors used the wrong criteria and incorrectly determined the types of drivers. (3) Very often, authors associate the diturbance of the magnetosphere-ionosphere-atmosphere system caused by a complex driver (by a sequence of single drivers) with one of the drivers, ignoring the complex nature. For example, magnetic storm are often caused by compression region Sheath in front of the interplanetary CME (ICME), but the authors consider this events as so-called "CME-induced" storm, not "Sheath-induced" storm. (4) Finally, there is a "lost driver" of magnetospheric disturbances: some authors simply do not consider the compression region Sheath before ICME if there is no interplanetary shock (IS) before Sheath, although this type of driver, "Sheath without IS", generates about 10% of moderate and strong magnetic storms.

## 1. Introduction

The pioneering studies in the 60s and 70s (Dungey, 1961; Fairfield et al., 1966; Rostoker et al., 1967; Russell et al., 1974; Burton et al., 1975) showed that disturbances in the magnetosphere are mainly associated with the appearance of the southward ($Bz < 0$) component of the interplanetary magnetic field (IMF). IMF lies in the ecliptic plane under steady interplanetary conditions and substantial $Bz < 0$ is observed only in disturbed types of solar wind (SW) such as corotating interaction regions (CIR) between slow and fast SW streams and interplanetary coronal mass ejections (ICME) and compression regions Sheaths in front of fast ICMEs (see reviews Tsurutani etal., 1997; Gonzalez et al., 1999; Yermolaev et al., 2005). There are many studies which shows different magnetosphere response on various types of solar wind, even for close values of IMF $Bz$ (Eselevich et al., 1993; Huttunen et al., 2002, 2006; Huttunen and Koskinen, 2004; Borovsky and Denton, 2006; Pulkkinen et al., 2007; Yermolaev et al., 2007a; Plotnikov and Barkova, 2007; Longden et al., 2008; Turner et al., 2009; Guo et al., 2011; Nikolaeva et al., 2013, 2014, 2015a, 2015b; Yermolaev et al., 2010a, 2010b, 2012, 2014,2015; Borovsky et al., 2016; Lockwood et al., 2016; Boroyev and Vasiliev, 2018; Despirak et al., 2019 ). Currently, this approach seems very promising, since it allows one to discover new physical connections in solar-terrestrial physics. There is currently a steady upward trend in the number of studies in which some magnetospheric, ionospheric, and atmospheric processes are compared with some specific types of solar wind. However, most researchers are not specialists in the solar wind phenomena and make mistakes in identifying interplanetary drivers, which often lead to incorrect conclusions. The most common errors are associated with incorrect criteria for identifying the types of solar wind, either by the authors of the erroneous work, or by the authors of those data sources that are used by other researchers. Typical examples of such errors were considered in detail in our works (Yermolaev et al., 2017; Lodkina et al., 2018) and will not be considered in this article.

In this paper, we consider two other incorrect approaches that lead to erroneous conclusions about the relationship of interplanetary drivers and magnetospheric disturbances. Firstly, authors associate the perturbation of the magnetosphere-ionosphere system caused by a complex driver (by a sequence of single drivers) with one of the drivers, ignoring the complex nature. For example, a magnetic storm is often caused by a compression region Sheath in front of an interplanetary CME (ICME), but the authors consider the ICME to be a cause of disturbance, not Sheath. Secondly, there is a "lost driver" of magnetospheric disturbances: some researchers simply do not consider the Sheath compression region before ICME if there is no interplanetary shock (IS) before Sheath, although this type of driver, "Sheath without IS", can generate moderate and strong magnetic storms.

The structure of this paper is as follows. Section 2 describes data and methods used. Section 3 presents the results of the Sheath measurements. Section 4 discusses and summarizes the results.

## 2. Data and Methods

In this paper, we use the following data and methods.

The basis of our investigation is the 1-h interplanetary plasma and magnetic field measurements and magnetospheric data of OMNI database (http://omniweb.gsfc.nasa.gov King and Papitashvili, 2004).

Using threshold criteria for key parameters of SW and IMF, we identified corresponding large-scale types of SW for every 1-h point of the archive during 1976–2018 (see paper by Yermolaev et al., 2009, and site ftp://ftp.iki.rssi.ru/pub/omni/). Our identification of SW types is based on methods similar to ones described in many papers and basically agrees with the results of other authors, but in contrast with other similar studies, we used a general set of threshold criteria for all SW types and made the identification for each 1-h point. To analyze the magnetosphere response on the change of interplanetary conditions, we select the following disturbed types of solar wind, i.e., corotating interaction region (CIR), two types of ICMEs (MC and Ejecta), two types of Sheath (SHMC and SHEJ), and IS forward shocks.

We use the double superposed epoch analysis (DSEA) method with 2 reference time instants at the ends of interval (Yermolaev et al., 2010a). This method involves re-scaling (proportionally increasing/decreasing time between points) the duration of the interval for all SW types in such a manner that, respectively, times of first and last points of all intervals of a selected type coincide.

A magnetic storm is considered to be associated with a solar wind phenomena if the moment of a minimum in the Dst index for the storm falls within the time interval of the SW event or is observed during 1–2 h after this phenomena (2 h correspond to the average time delay between the Dst peak of an intense magnetic storm and the associated peak in the southward IMF Bz component (Gonzalez and Echer, 2005, Yermolaev et al., 2007a,b)).

## 3. Results

In our papers (Yermolaev et al., 2015; 2017) using the double superposed epoch analysis method, we studied the average behavior of interplanetary and magnetospheric parameters for the eight usual sequences of SW phenomena: (1) SW/CIR/SW, (2) SW/IS/CIR/SW, (3) SW/Ejecta/SW, (4) SW/Sheath/Ejecta/SW, (5) SW/IS/Sheath/Ejecta/SW, (6) SW/MC/SW, (7) SW/Sheath/MC/SW, and (8) SW/IS/Sheath/MC/SW (where SW means undisturbed solar wind and IS means interplanetary shock) for 1976–2000 and showed that the average temporal profiles of magnetospheric indices have the maxima in intervals between Sheath end and ICME beginning. In particular, the average temporal profiles of measured Dst and density-corrected Dst* indices, which mainly reflect the behavior of the ring current, are divided into two parts (see panels in the first and third rows of Fig.1), i.e., (1) the drop in Dst and Dst* indices is observed in the Sheath (with minima of –50 nT in the early hours in the MC and –35 nT in the Ejecta, respectively, and the Dst* index is systematically 5–10 nT lower than Dst) and (2) the slight increase in Dst and Dst* indices in the MC and Ejecta. For the MC and Ejecta with Sheath and IS, in general, the picture is identical for MC and Ejecta with Sheath and without IS, the only difference being that the Dst and Dst* minima are –70 and –50 nT. The fact that the corrected Dst* index in the Sheath is systematically lower than the measured Dst index is associated with higher values of density and pressure in the Sheath regions compared to the MC and Ejecta.

The panels of the second and fourth rows of Fig. 1 show the time distributions for the Sheath and Ejecta/MC intervals, respectively, of the number of the following events: the onsets of storms with Dst < –50 nT (blue columns) and the Dst minima (red columns). Though the blue and red columns in the figure are shifted with respect to each other for clarity, they were calculated in the 5 identical subintervals. These data show that a great number of magnetic storms began at the beginning of Sheath, and the maximum number of Dst index minima (the maxima of magnetic storms) fell at the end of Sheath to the beginning of Ejecta/MC.

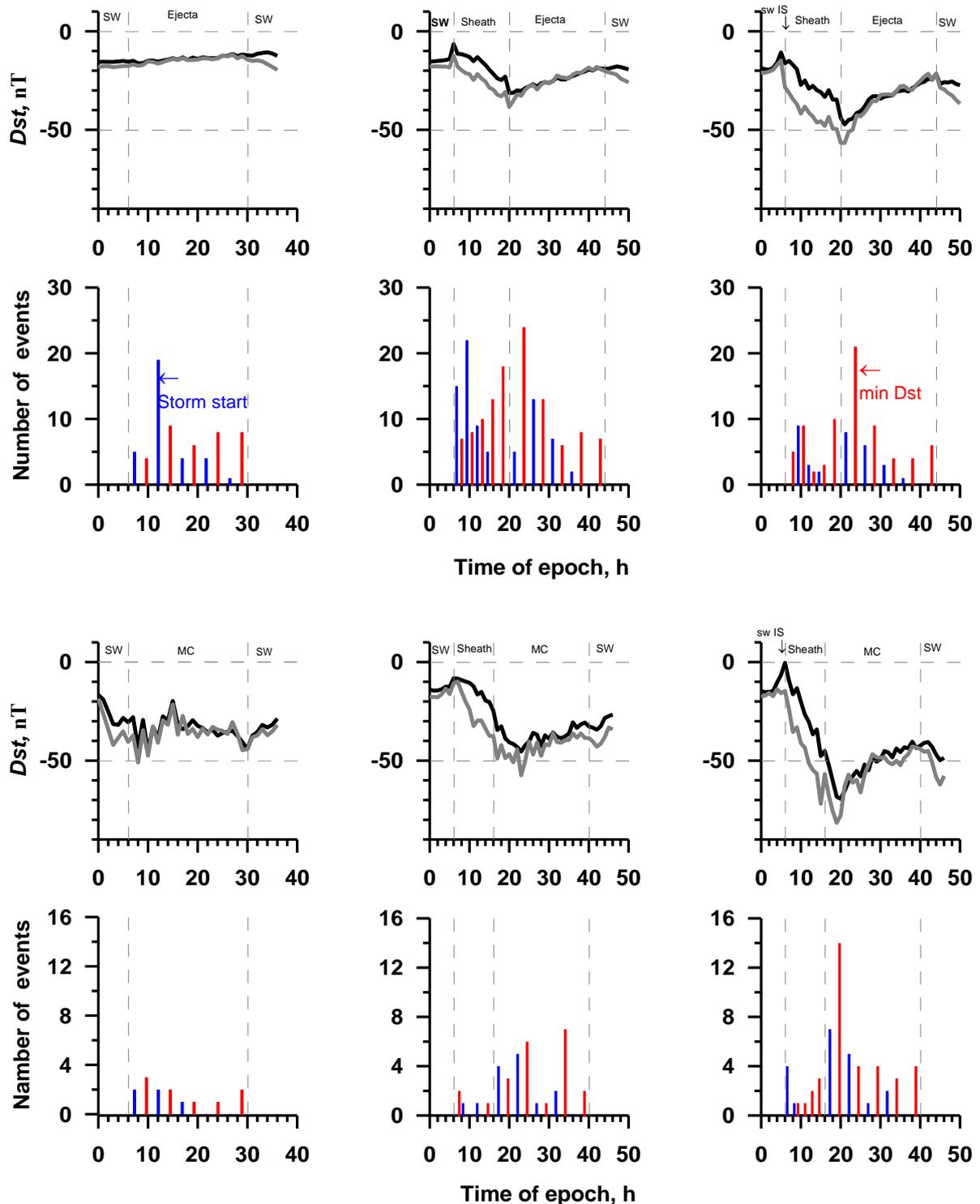

**Fig. 1.** *Temporal profile of Dst (black) and Dst* (gray) indices for six different sequences of solar wind phenomena. Vertical dashed lines indicate (from right to left): 1. last point of the Ejecta/MC intervals; 2. first point of the Ejecta/MC intervals; 3. (in the presence of Sheath) first point of the Sheath intervals. Panels of second and fourth rows show the distributions, in Sheath or Ejecta/MC time interval, number of beginnings of storms (blue columns) and number of maxima (Dst index minima) of storms (red columns).*

The table and figures 2-5 allow one to compare the Sheath characteristics in 4 variants of the sequence of SW types: IS/Sheath/Ejecta, Sheath/Ejecta, IS/Sheath/MC and Sheath/MC.

The number of Sheath events before Ejecta without IS (432) slightly exceeds the number of events with IS (381), and the number of Sheath before MC with IS (152) significantly exceeds the number of events without IS (28). Although the average values for many parameters of the Table turned out to be close in magnitude to the standard deviations, the statistical error

(standard deviation divided by the square root of the number of points) for some of them turned out to be small, and in this case, the differences in the mean values for different types of Sheath can be considered statistically reliable. In particular, the data in the Table show that the average duration of Sheath events before Ejecta is longer than before MC. The number of magnetic storms generated by Sheath before Ejecta with and without IS is almost the same (61 and 59), and for the MC the difference is more significant (24 and 3), but it was obtained with small general statistics of the MC compared with Ejecta.

**Table.** Mean values and standard deviations of parameters for 4 types of Sheath.

|  | S/Sheath/Ejecta | Sheath/Ejecta | IS/Sheath/MC | Sheath/MC |
|---|---|---|---|---|
| Number of events | 381 | 432 | 152 | 28 |
| Duration of events, h | 16.4±9.6 | 14.0±8.8 | 12.4±6.1 | 12.8±9.5 |
| Number of magnetic storms | 61 | 59 | 24 | 3 |
| $V$, km/s | 460±108 | 439±95 | 497±141 | 433±97 |
| $T$ ($10^5$), K° | 1.90±1.78 | 1.70±1.41 | 2.56±3.62 | 1.69±1.69 |
| $T/Texp$ | 2.18±1.24 | 2.29±1.23 | 2.21±1.67 | 2.28±1.29 |
| $N$, cm$^3$ | 12.4±9.5 | 9.7±6.4 | 16.1±11.3 | 13.5±8.3 |
| $B$, nT | 9.9±4.7 | 8.2±3.6 | 13.6±7.8 | 10.0±5.1 |
| $Kp*10$ | 33±16 | 29±15 | 42.8±18.9 | 31±16 |
| $Dst$, nT | -19±36 | -18±27 | -24±54 | -17±27 |
| $Dst*$, nT | -28±39 | -23±29 | -37±55 | -26±26 |
| $AE$, nT | 329±286 | 278±250 | 458±394 | 303±306 |

Figures 2-5 show the temporal profiles of interplanetary parameters and magnetospheric indices for 4 sequences: (1) SW / IS / Sheath / Ejecta, (2) SW / Sheath / Ejecta, (3) SW / IS / Sheath / MC and (4) SW / Sheath / MC. Each of the four figures contains 10 panels, which depict the average temporal profiles of the analyzed parameters obtained by DSEA method for Sheath regions (for the solar wind before Sheath from 0 to 5 points and ICME after Sheath from 20 to 25 points, the simple method of superposed epoch analysis with reference points was used at the ends of the Sheath interval):

a: the ratio of thermal and magnetic pressure β, thermal pressure Pt, the relative density of alpha particles Na / Np,
b: proton temperature T * $10^{-5}$ K and the ratio of measured and expected temperatures T / Texp,
c: the angles of the velocity vector Phi, Theta,
d: the component of the IMF Bz, the component of the electric field Ey,
e: Dst and Dst * indices,
f: the magnitude of the IMF B and the dynamic pressure Pd,
g: components of the IMF Bx, By,
h: sound and Alfen speeds Vs and Va,
i: the ion density N and Kp index,
j: the plasma velocity V, index AE.

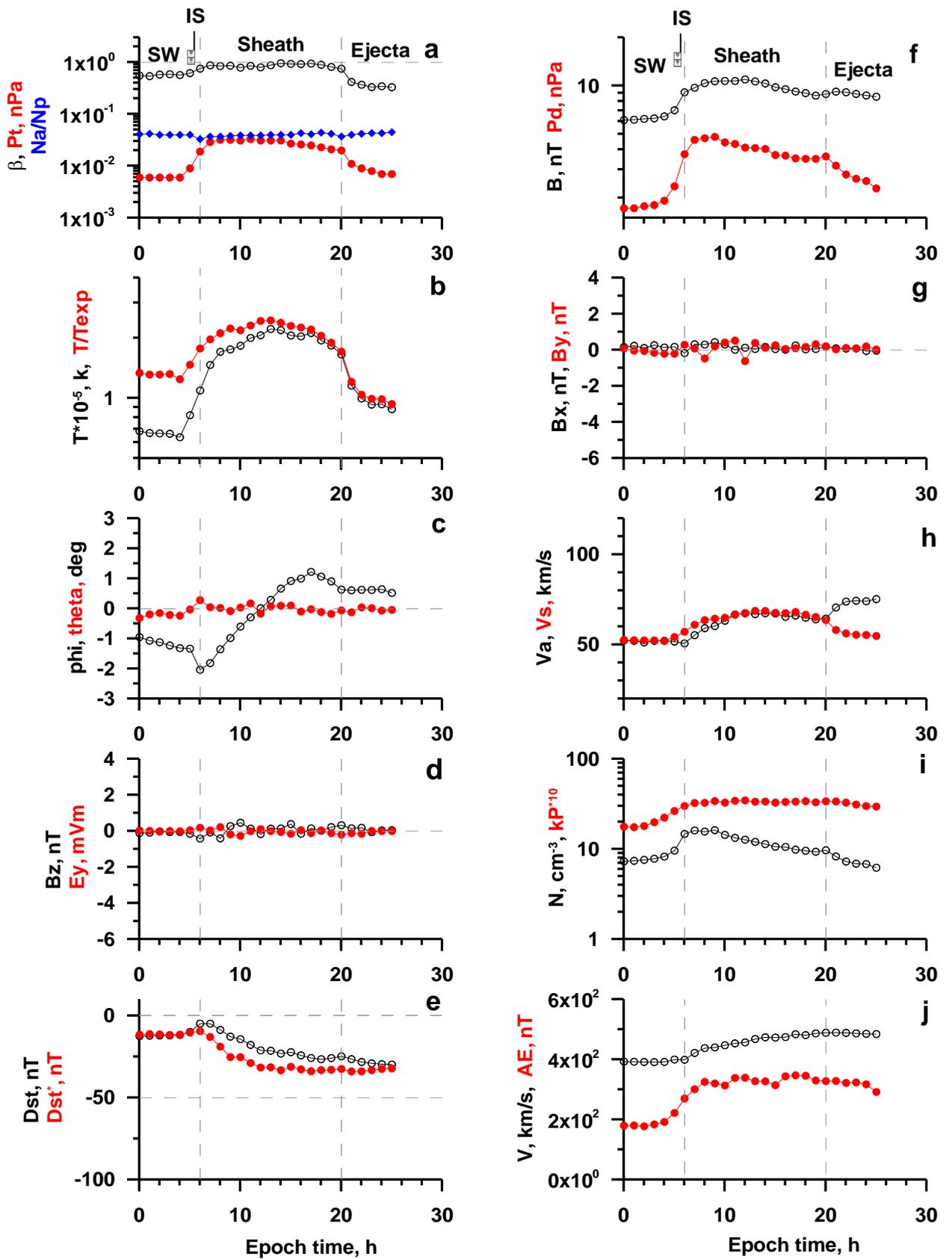

**Fig.2** *The temporal profiles of the solar wind parameters and magnetospheric indices for the IS/Sheath/Ejecta sequence obtained using the SEA and DSEA methods: from 0 to 5 and 20–25 points, SEA was used without re-scaling; from 6–19 points, DSEA was used with re-scaling up to 14 points*

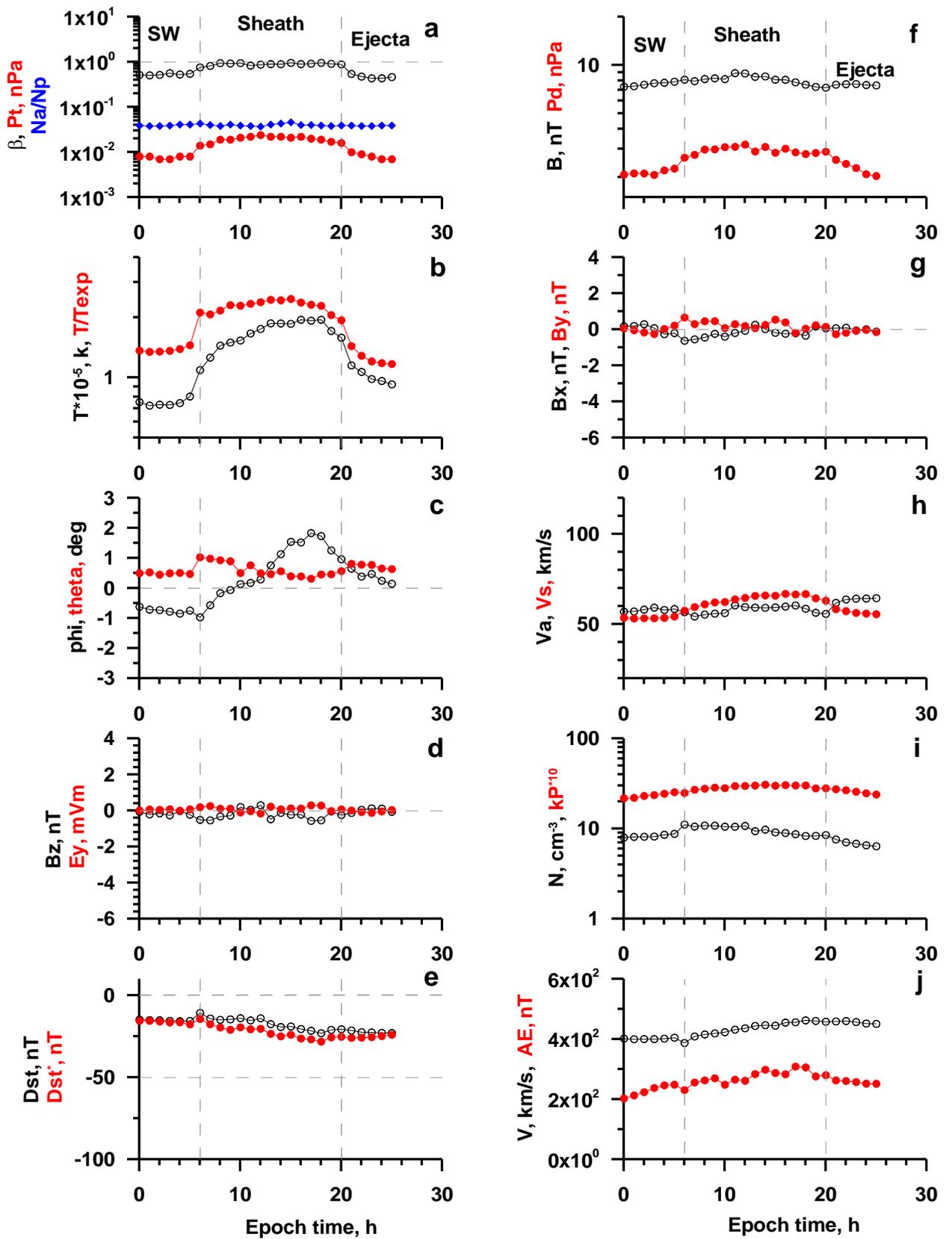

**Fig.3.** The *same as in Fig. 2 for the Sheath/Ejecta sequence*

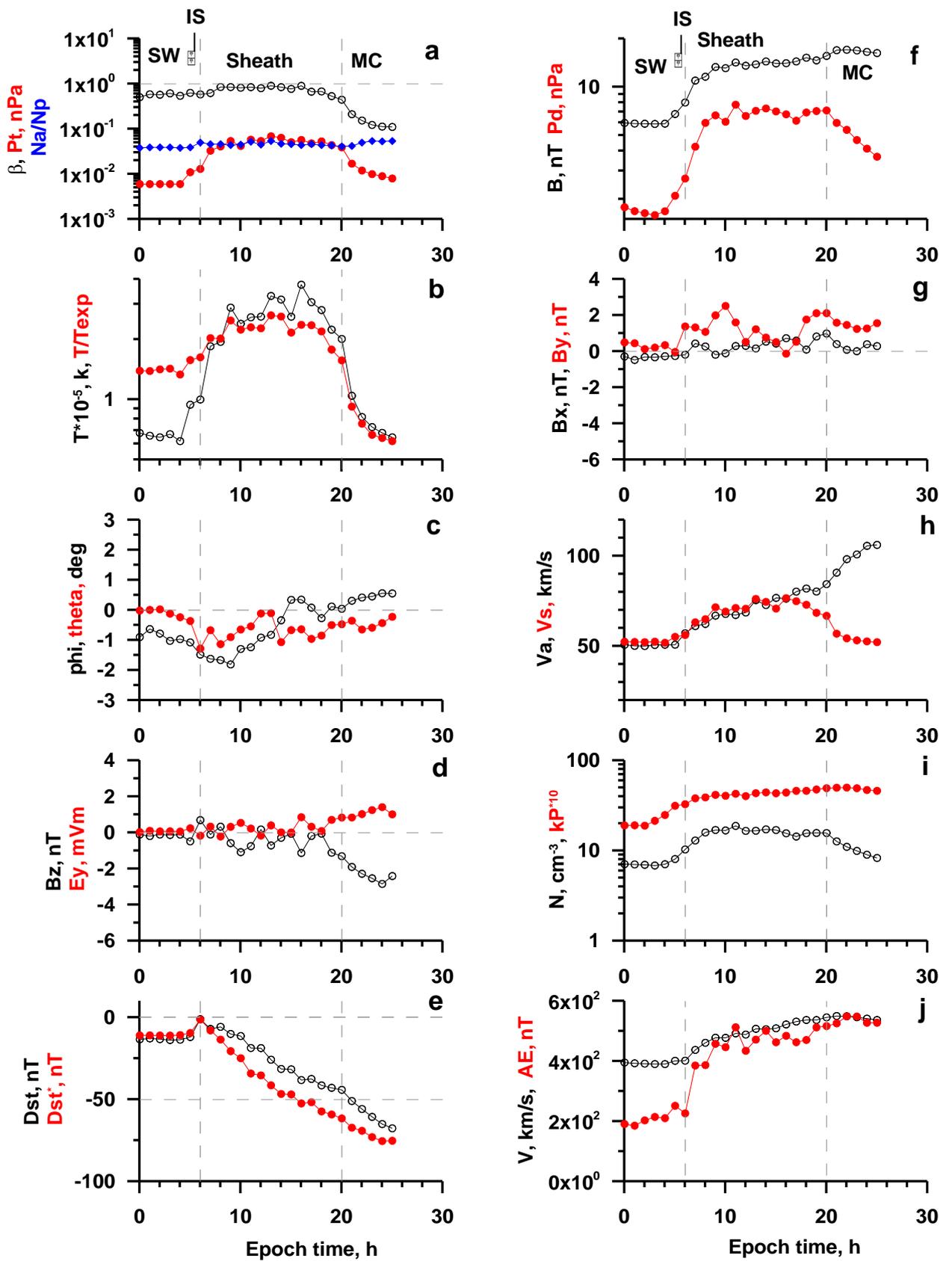

**Fig.4.** The *same as in Fig. 2 for the IS/Sheath/MC sequence.*

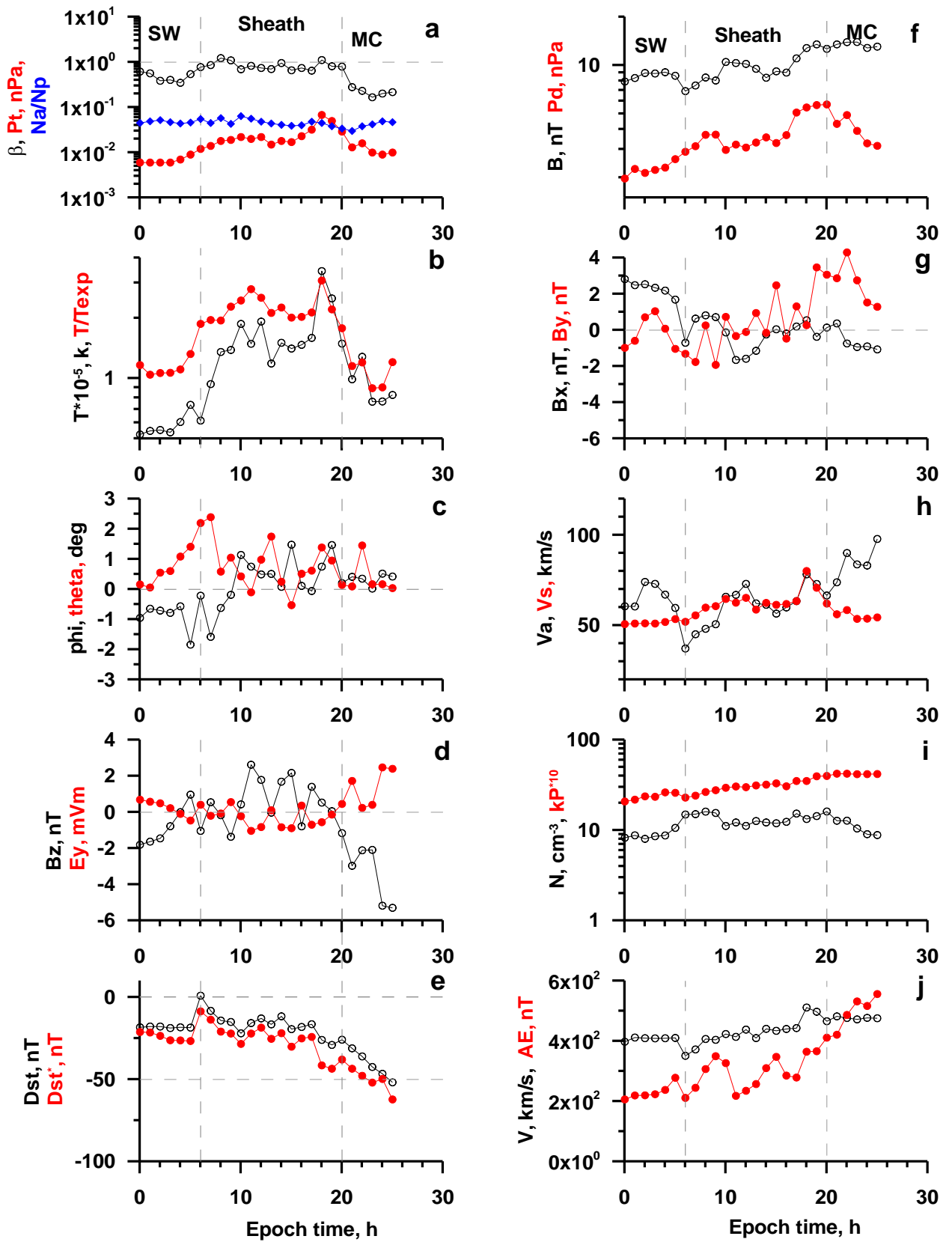

**Fig.5.** The *same as in Fig. 2 for the Sheath/MC sequence*.

For "Sheath with IS" events followed by Ejecta, the values of such parameters as the magnitude of the magnetic field B, thermal and dynamic pressure Pt, Pd, plasma velocity V, proton temperature T and T / Txp, Dst and Dst * indices are larger than for "Sheath without IS" (Fig.2 and 3). This is mainly due to a sharper increase in these parameters if the Sheath region begins with IS, since in the subsequent time after 2-3 hours after the start of the Sheath region, these parameters change in a similar way. The situation is similar for "Sheath with IS" and "Sheath without IS " events followed by MC (Figs. 4 and 5). For Sheath events with subsequent MS, the values of the parameters B, Pt, Pd, V, T, T / Texp, Dst, Dst * are higher than with the subsequent Eject.

## 4. Discussion and conclusions

Many studies investigated so-called "CME-induced" storms (or other types of magnetospheric disturbances) as an independent type of storms. In our opinion, there are no CME-induced disturbances, but there are Sheath-induced and MC/Ejecta-induced disturbances, as well as multi-step disturbances, which are excited by a sequence of Sheath/MC or Sheath/Ejecta events. The presented data indicate that the CME-induced disturbances of the magnetosphere can represent the response to absolutely different interplanetary drivers or their successive impact. The region "Sheath without shock" is observed before ICME almost as often as the Sheath region with IS, is sufficiently geoeffective and is the driver of about 10% of all storms. These drivers have different physical natures, possess different efficiencies of the impact on the magnetosphere and may lead to the implementation of different mechanisms of this impact.

   The following experimental facts should be mentioned. (1) The average magnitude of IMF B in Sheaths is higher than B in Ejecta and is close to B in MCs (Yermolaev et al., 2015). (2) The efficiency of magnetic storm generation is 50% higher for Sheath than for ICME (MC and Ejecta) (Nikolaeva et al., 2013, 2015; Dremukhina et al., 2018, 2019), i.e. at identical southward components of the interplanetary magnetic field, the magnetic storms are generated ~1.5 times more strongly by Sheaths than by ICMEs.

Thus, it is possible to conclude that, in our opinion, the contribution of compression regions Sheath (including lost driver: Sheath without shock) in the generation of storms is often not taken into account and their role is often underestimated, and this erroneous approach often results in incorrect conclusions during studying the solar-terrestrial links.

ACKNOWLEDGMENTS
The authors express their gratitude to the developers of the OMNI database (http://omniweb.gsfc.nasa.gov) for the opportunity to use it in the work. YY is grateful to the SCOSTEP's "Variability of the Sun and Its Terrestrial Impact" (VarSITI) program for support of participation in the Closing Symposium, June 10-14, 2019, Sofia, Bulgaria. The work is supported by the RFFI grant 19-02-00177a.